\documentclass[aps,prl,preprint,superscriptaddress]{revtex4-1}
\usepackage[english]{babel}
\usepackage{graphicx}
\usepackage{bm}
\usepackage[latin1]{inputenc}
\usepackage{amsfonts}
\usepackage{amssymb}
\usepackage{amsmath}
\usepackage{hyperref}
\hypersetup{colorlinks=true}

\newcommand{\de}{\mathrm{d}}
\newcommand{\Tw}{T\!w}
\newcommand{\Wr}{W\!r}
\newcommand{\wwr}{wr}

\newcommand{\Lk}{n}
\begin{document}
\title{Multi-plectoneme phase of double-stranded DNA under torsion}
\author{Marc Emanuel}
\affiliation{Instituut Lorentz voor de theoretische %
natuurkunde, Universiteit Leiden,%
P.O. Box 9506, NL-2300 RA Leiden, The Netherlands}
\affiliation{Institute of Complex Systems II, Forschungszentrum J\"ulich, J\"ulich 52425, Germany}
\author{Giovanni Lanzani}
\author{Helmut Schiessel}
\affiliation{Instituut Lorentz voor de theoretische %
natuurkunde, Universiteit Leiden,%
P.O. Box 9506, NL-2300 RA Leiden, The Netherlands}
\pacs{64.70.km,87.10.Ca,87.15.ad}
\begin{abstract}
  We use the worm-like chain model to study
  supercoiling of DNA under tension and torque. The model reproduces
  experimental data for a much broader range of forces, salt concentrations and
  contour lengths than previous approaches. Our theory shows, for the first
  time, how the behavior of the system is controlled by a multi-plectoneme
  phase in a wide range of parameters. This phase does not only affect
  turn-extension curves but also leads to a non-constant torque in the
  plectonemic phase. Shortcomings from previous models and inconsistencies
  between experimental data are resolved in our theory without the need of
  adjustable parameters.
\end{abstract}
\maketitle
The DNA contained in every cell of all higher organisms is hundred
times longer than the cell diameter: to fit inside it has to fold. This is a
challenging problem since DNA is a semi-flexible polymer, making it hard
to confine it in small spaces. On the other hand, local unfolding of
DNA has to be efficient, as it plays a key role in the
transcription and replication of the genome. Unfolding is achieved by
stretching and twisting the molecule: unraveling how DNA reacts to
them is crucial to understand cellular activities.

The relevant mechanical properties of DNA have been studied with
single molecule techniques, where individual molecules are stretched
and/or twisted under physiological conditions. The stretching and
bending elasticity, in the absence of twisting, has been investigated
through measurements of the force-extension relation of
DNA~\cite{Smith:1992} and theories based on the worm-like chain
(WLC) model have successfully explained the experimental
results~\cite{Marko:1995a}~\cite{Odijk:1995}. The WLC model~\cite{Doi:1988}
is a coarse-grained approximation where the particular sequence of
basepairs (bp) is hidden by treating the DNA as a homogeneous semiflexible polymer.

\begin{figure}[htb]
  \centering
  \includegraphics[width=7.9cm]{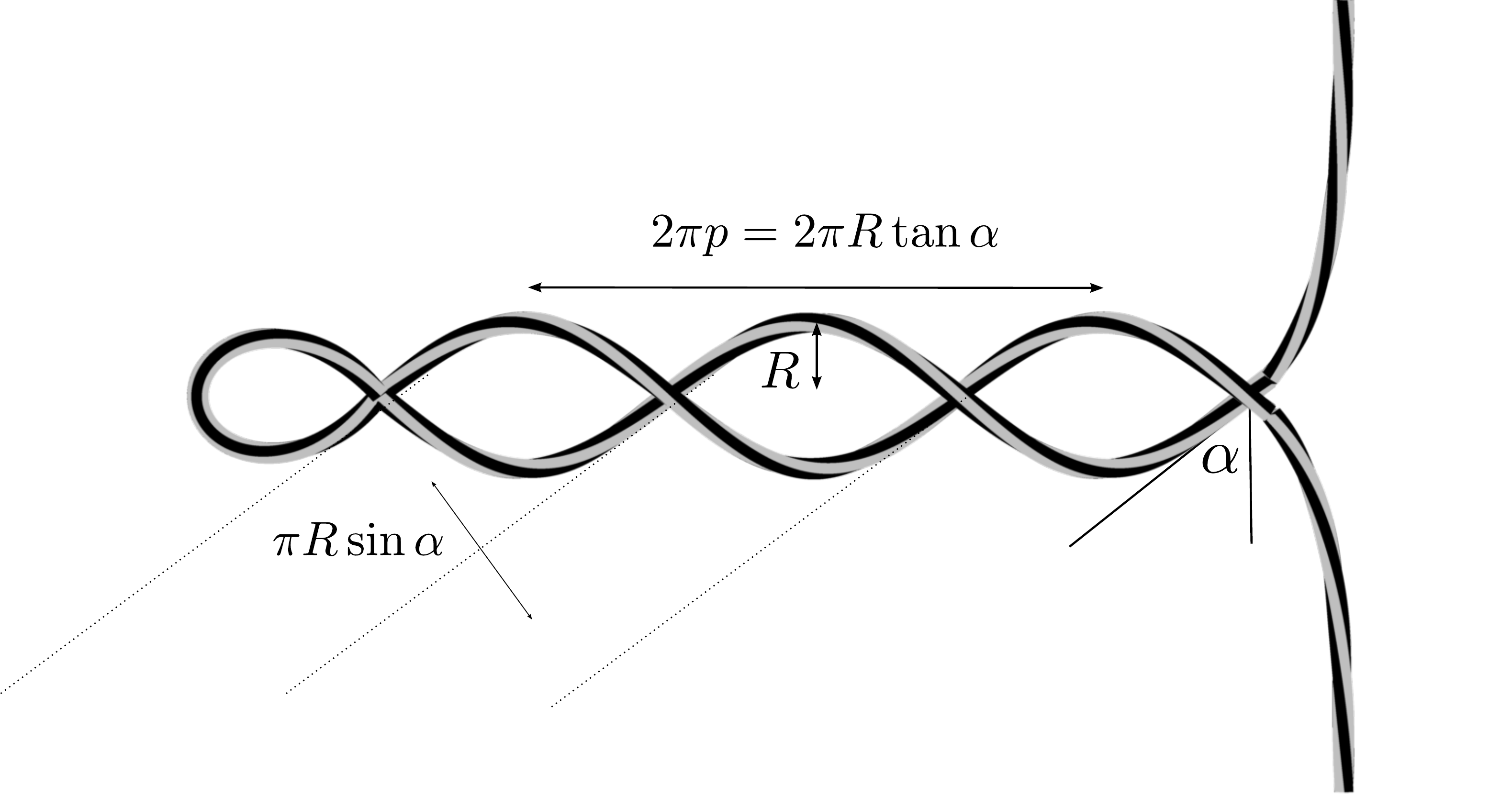}
  \caption{Geometry of the plectoneme.}
  \label{fig:new_homoclinic}
\end{figure}
To understand stretched DNA under torsional
stress~\cite{Strick:1996,Forth:2008,Mosconi:2009}
several models based on the WLC framework were proposed. However, they
were either purely mechanical~\cite{Purohit:2008}, involved
non-linear elasticity~\cite{Smith:2008},
phenomenological~\cite{Marko:2007}, aimed only at
a certain region of the experimental data~\cite{Clauvelin:2008} or had to invoke
a non-canonically reduction of the DNA charge density~\cite{Maffeo:2010}. The outcome of
the experiments still remains poorly understood.

\begin{figure}[htb]
  \centering
  \includegraphics[width=7.0cm]{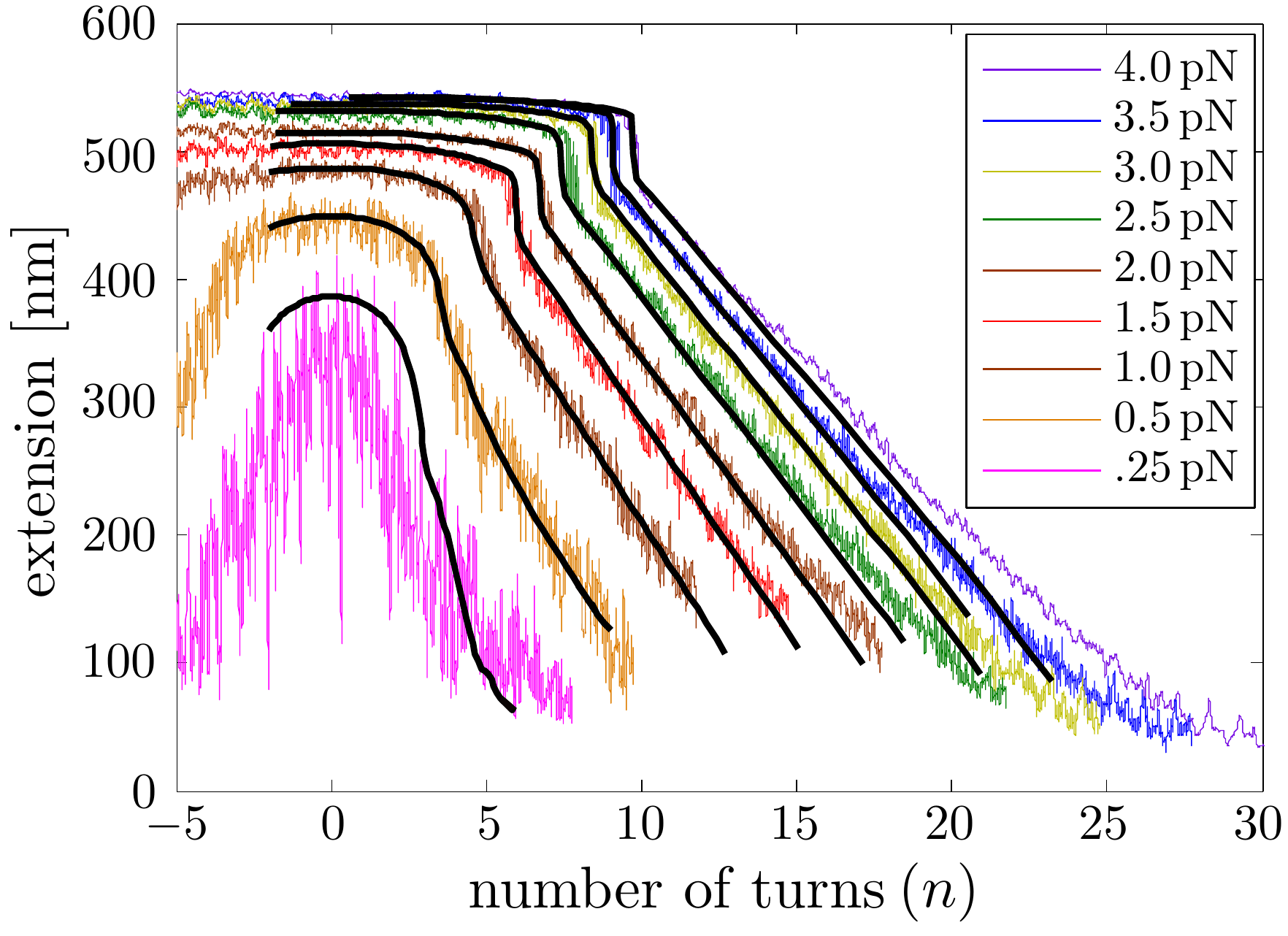}
  \caption{The turn-extension plots of a $600\,$nm DNA chain at $320\,$mM ionic
      strength for various tensions between $0.25\,$ and $4\,$pN. Comparison
      between theory and experiments. Experimental data from~\cite{Maffeo:2010,*Brutzer:2010}.}
  \label{fig:data}
\end{figure}

\begin{figure}[htb]
  \centering
  \includegraphics[width=7.0cm]{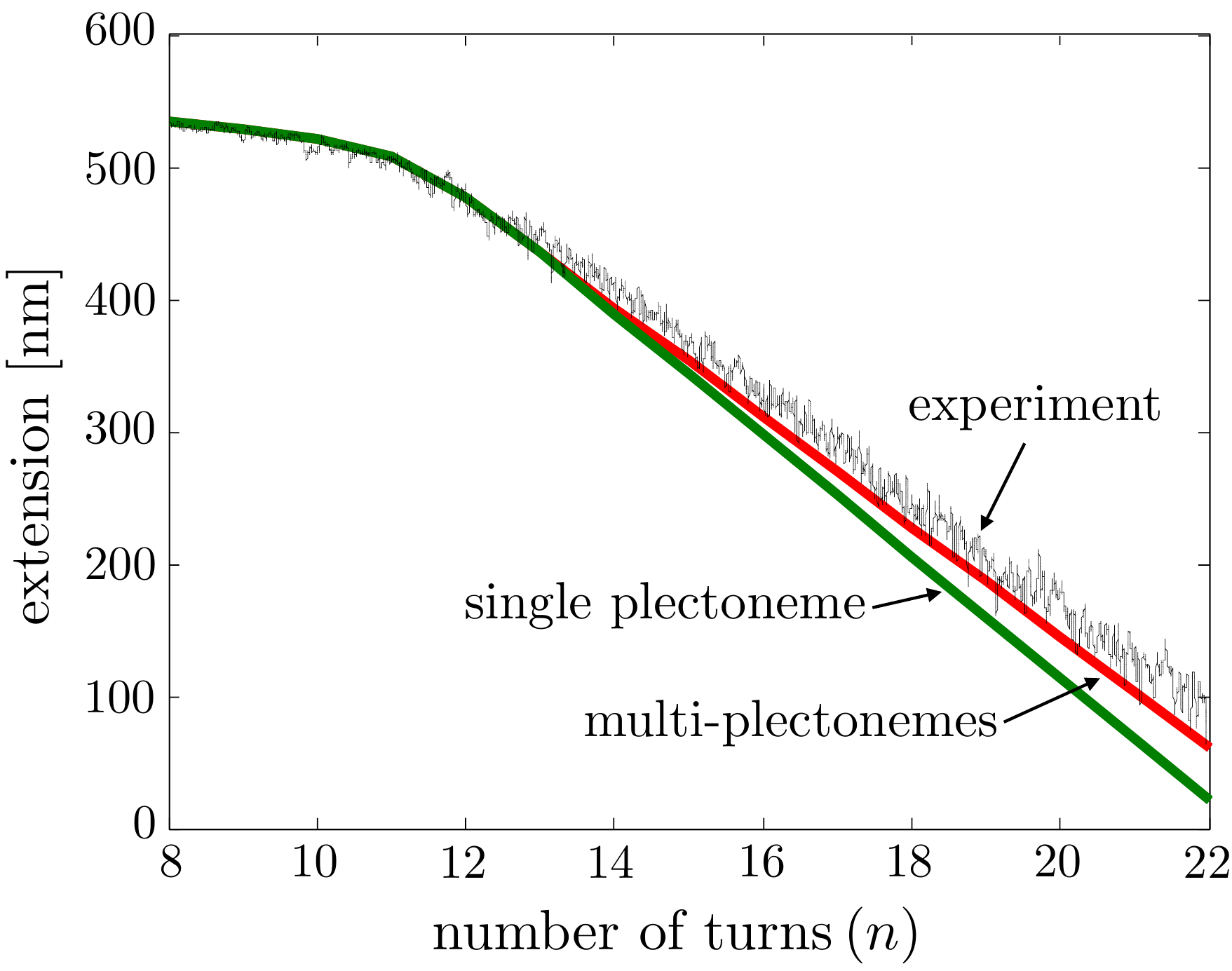}
  \caption{The results of the theory with and without the possibility to form
      more than one plectoneme are presented alongside the experimental results
      ($3\,$pN, $20\,$mM, Experimental data from~\cite{Maffeo:2010,*Brutzer:2010}.}
  \label{fig:singmult}
\end{figure}

\begin{figure}[htb]
  \centering
  \includegraphics[width=7.0cm]{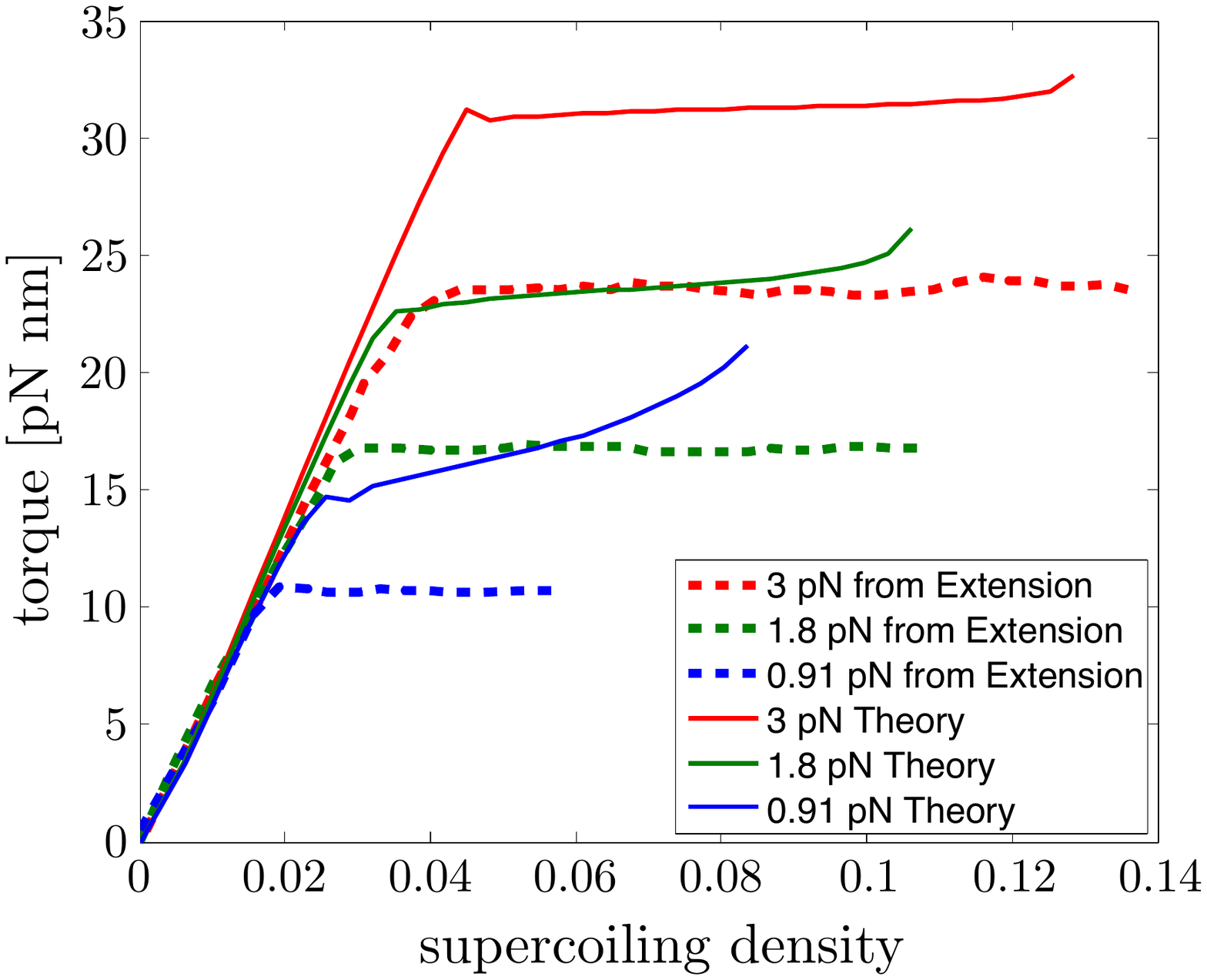}
  \caption{The supercoiling density-torque plots of a $5600\,$nm DNA chain at $100\,$mM ionic
      strength for $0.91\,$pN, $1.8\,$pN and $3\,$pN tension. Comparison
      between theory and torques inferred from extension measurements. Data taken from~\cite{Mosconi:2009}}
  \label{fig:torquem}
\end{figure}

In this Letter we present a theory, based on the WLC model, without
any of the aforementioned shortcomings.
The results describe experimental data accurately (see
Fig.~\ref{fig:data} for an example). Up to now it has been assumed that
under high twist DNA reduces its torque through the formation of a \emph{single} plectoneme,
see Fig.~\ref{fig:new_homoclinic}. We show here for the first time how thermal fluctuations
lead to a \emph{multiple} plectoneme phase instead. We demonstrate its impact on
the turn-extension slope, see Fig.~\ref{fig:singmult}, and on the torque, see Fig.~\ref{fig:torquem}.

In the experiments the DNA is anchored at one end to a surface and at the other
end to a magnetic~\cite{Mosconi:2009} or optical~\cite{Forth:2008} tweezer. 
This allows to control the tension and the linking
number (\(\Lk\)), the number of turns inserted, at the same time. Increasing
\(\Lk\) at constant tension yields turn-extension plots like the ones shown in
Fig.~\ref{fig:data}. Initially most of \(\Lk\) goes into twist (\(\Tw\)) of the
molecule while the end-to-end distance remains approximately constant.  Then a
transition --- dependent on the salt concentration, the DNA length and the
applied force and often accompanied by a jump~\cite{Forth:2008} --- signals the
formation of a plectoneme. From this point onward all the additional \(\Lk\) is
stored inside the growing plectoneme as writhe (\(\Wr\)), a quantity related to
the path of the polymer. Writhe, twist and linking number are
connected through \emph{White's relation}~\cite{Calugareanu:1959,*White:1969}, \(
\Lk=\Tw+\Wr\).

We assume that the legs and the end loop form a
homoclinic solution, characterized by the parameter \(t \in [0,1]\), as
described in~\cite{Nizette:1999}.  \(t=0\) corresponds to a
straight rod, and \(t=1\) to the homoclinic loop. Inside the homoclinic solutions, at the
point of (non-zero) minimum distance of symmetric points, we insert a
plectoneme. Such a minimum exists in the range \(0.804\gtrsim t>1 \);
half this distance sets the plectoneme radius
\(R(t)\). The homoclinic solution stores some fixed amount of writhe,
\(\Wr_l= 2 (\arcsin t) /\pi\). Moreover its bending and potential energy
sum up to \(E_l= 8 F \lambda t\). Here \( F\) is the tension,
\(P_b=A/k_B T\) the DNA bending persistence length and \(\lambda= \sqrt{A/F}\).

On the other hand, the plectoneme has a bending energy density \(e_b = A \cos^4
\alpha /2 R^2(t)\) and a contribution to the potential energy density of \(f \)
where \(\alpha\) denotes the angle of the
plectoneme, Fig.~\ref{fig:new_homoclinic}. When properly accounting for the
presence of the end loop, the plectoneme writhe density is exactly given by
\(\wwr_p(t)
= \sin \alpha \cos \alpha / 2 \pi R(t)\). The electrostatic interaction in
the plectoneme has a free energy density \(f_{\mathrm{el,0}}\)
described by Ubbink and Odijk~\cite{Ubbink:1999}. The effective charge density
in this contribution is calculated on the basis of a charge density of 2
charges/\(0.34\,\)nm with the method described in~\cite{Philip:1970}.

The zero-temperature energy density of a DNA chain with contour \( L_c \) containing
$m$ plectonemes of total relative length $l_p \equiv L_p / L_c$ is given by
\begin{align}
\notag  e_0 (m,l_p) &= l_p e_b + m \frac{E_l}{L_c}+
l_p(F+f_{\mathrm{el,0}})\\&+
2 \pi^2 P_C k_B T(\frac{\Lk}{L_c}-m\frac{\Wr_l}{L_c}- l_p\wwr_p)^2
\label{eq:energy}
\end{align}
where the last term is the twist contribution to the energy, and
\(P_C\) is the torsional persistence length. However, to
properly account for the experimental situation the zero-temperature
analysis is not sufficient.

Thermal fluctuations lead to three different contributions: the first affects
the DNA legs, the second acts within plectonemes and the last is related to the
number of plectonemes and their position and length distribution.  Outside the
plectonemes, thermal fluctuations modify the shape of the DNA; for a given
torque \( \tau\) they shorten the DNA end-to-end distance by a factor
\(\rho(F,\tau)\)~\cite{Moroz:1998} and give origin to thermal writhe that
lowers the twist energy density, which can be expressed as a renormalized
torsional persistence length \( P_C^{\mathrm{eff}}(\lambda)
\)~\cite{Moroz:1998}.

Inside the plectonemes, following Ref.~\cite{Ubbink:1999} we consider
fluctuations in two directions.  We denote by \(\sigma_r\) (\( \sigma_p \)) the
standard deviation in the radial (pitch) direction.  The fluctuations in the
pitch direction are set by the geometry: \(\sigma_p = \pi R\sin \alpha\),
see Fig.~\ref{fig:new_homoclinic}, while in the other direction they are set by
the electrostatic repulsion.  Their contribution changes the electrostatic
energy, up to first order, to \(f_{\mathrm{el}} =
f_{\mathrm{el,0}} \times \exp (4 \kappa^2
\sigma_r^2)\)~\cite{Ubbink:1999}, where \(\kappa^{-1}\) is the Debije length.
The presence of twist couples the two directions of the fluctuations, an effect
not studied before~\footnote{Marc Emanuel, in preparation.}. This result in a new effective deflection
length: 
\begin{align}
 \bar{\lambda}&=2\frac{\lambda_r^3\lambda_p+\lambda_r^2\lambda_p^2+\lambda_r\lambda_p^3}{(\lambda_r+\lambda_p)(\lambda_r^2+\lambda_p^2)} 
\end{align}
where  \(\lambda_{r,p}
= (P_b \langle \sigma_{r,p}^2 \rangle)^{1/3} \) are the deflection lengths of confinement as defined in Ref.~\cite{Odijk:1983}. The resulting twist energy
density,
renormalized with \( P_C^{\mathrm{eff}}(\bar{\lambda}) \), depends on the
confinement. Since twist diffusion happens on a very short time scale, the
twist energy density in the legs and in the plectonemes should be the same.
This non-trivially couples the linking number density between legs and
plectonemes.

As a consequence of thermal fluctuations, the plectonemes' path  is
shortened by \(\rho_{\mathrm{str}}\), its bending energy density (\( e_b \)) and bare
writhe density (\( \wwr_b \))  are modified; the confinement of the polymer
contributes an additional \(f_{\mathrm{conf}}\) to the free energy~\cite{Marc:2012}. Together
with the change from \( f_{\mathrm{el},0} \) to \(
f_{\mathrm{el}} \), these modify \( e_0(m, l_p) \) to \( f(m, l_p) \).

In the infinite chain limit, as long as the number of plectonemes stays small,
the plectoneme parameters \( R,\,\alpha\) and \( \sigma_r\) become independent
of \( l_p \) since end loop contributions drop out.

The final contribution of the fluctuations comes from two
combinatorial factors in the partition function. They arise from the
number of ways the total length of the plectonemes can be distributed
between them, and the number of ways the plectonemes can be placed
along the DNA. Since the quantities involved are continuous we need to impose a
\(\xi\)-cutoff in our calculations which we assume in the following to
be given by the DNA helical repeat, $\xi=3.4\,$nm. Assuming hard-core
interactions between the plectonemes, this results in the partition
function
\begin{align}
  Z &= Z_0 + \sum_{m=1} \int \de L_p G(m,L_p) e^{-\beta f(m,l_p) L_c}\\
  G &= \frac{(\rho(L_c-L_p)-mL_{\mathrm{loop}})^m}{\xi^m m!}\frac{L_p^{m-1}}{\xi^{m-1}(m-1)!}.
\end{align}
Here \(Z_0\) is the partition function when \(m=0\) and \(L_{\mathrm{loop}}\) is the
length of a single end loop. The first factor of
\(G\) is the number of ways one can arrange \(m\) plectonemes along the DNA;
the second factor is the number of ways the length \(L_p\) can be distributed between
\(m\) plectonemes.

\begin{figure}[htb]
  \centering
  \includegraphics[width=7.0cm]{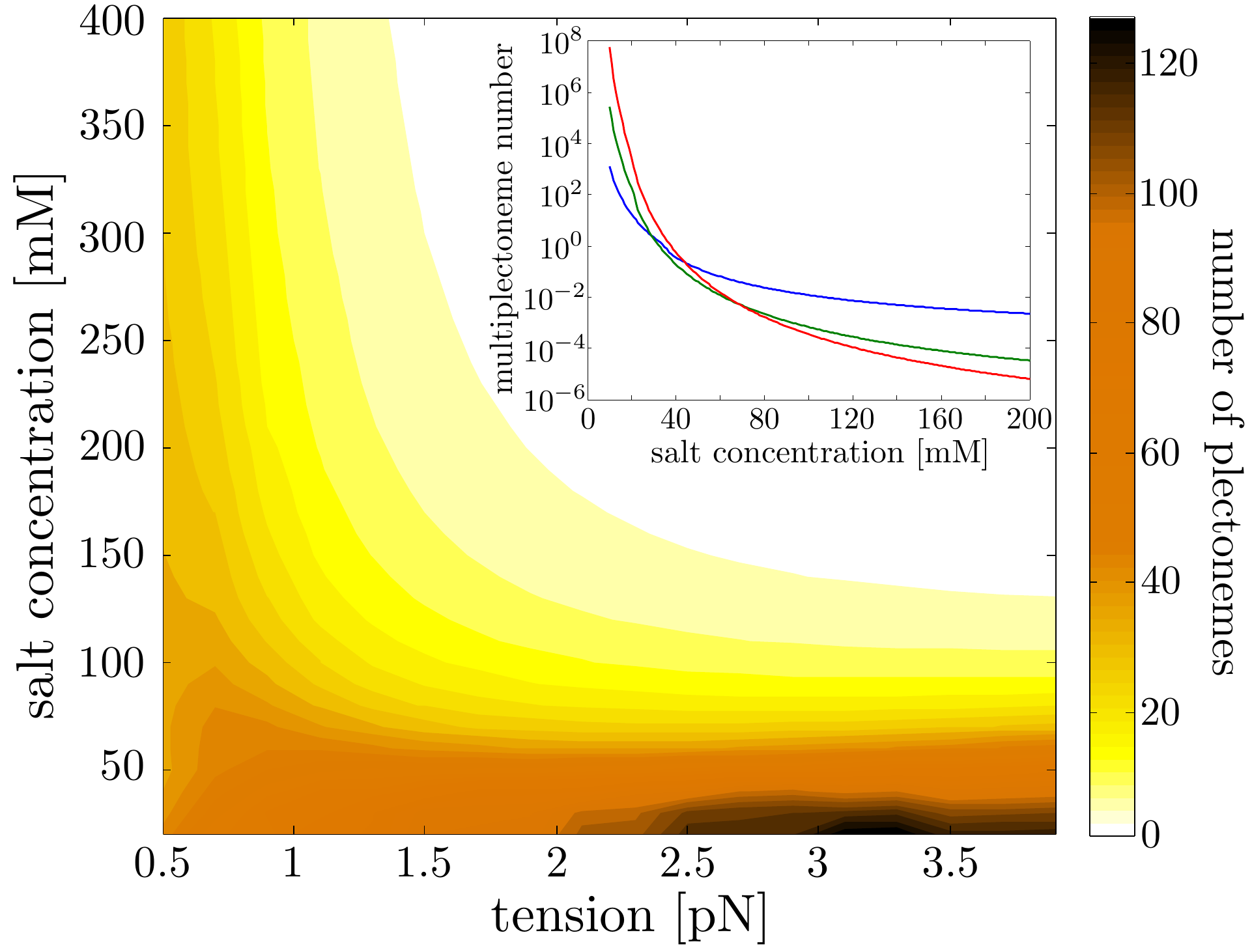}
  \caption{Phase diagram of the average number of plectonemes as a function of tension and salt concentration for a 7.2 $\mu$m long chain. Note the shift of
the maximum from low tension at high salt to high tension at low salt. The inset shows the multi-plectoneme parameter versus salt concentration for $1\,$pN (blue), $2\,$pN
(green) and $3\,$pN (red).}
  \label{fig:zeta}
\end{figure}

When \( \partial_n m L_{\mathrm{loop}} \ll \partial_n L_p \) the system is in
the single-plectoneme state.  On the other hand, for \( \partial_n m
L_{\mathrm{loop}} \approx \partial_n L_p \) the physics of increasing $n$ cannot be described by the notion of plectoneme length growth alone, but a full
multi-plectoneme approach is needed. To characterize these two states we introduce the multi-plectoneme parameter \( \eta \) as the
difference between the writhe efficiencies of loops and plectonemes
\begin{align}
    \eta \equiv \frac{E_l}{\Wr_l} - \frac{f_p}{\wwr_p}
    \label{eq:eta}
\end{align}
where \( f_p \) is the free energy density difference between plectoneme and
legs. \( \eta \) is important because it enters exponentially the
multi-plectoneme parameter \( \zeta \)
\begin{align}
    \zeta \equiv e^{- \Wr_l \eta} \left( \frac{\Wr_l/L_l}{\wwr_p}  \right)^2.
    \label{eq:zeta}
\end{align}
One can show that \( \zeta \ll 1 \) corresponds to a single-plectoneme state,
whereas
\( \zeta \approx 1  \) signals the multi-plectoneme phase. In
the inset of Fig~\ref{fig:zeta}, \( \zeta \) is displayed as a function of salt
concentration for different tensions.

Theories without the possibility of multiple plectonemes, typically predict,
for low salt concentrations, a too steep slope of the linear part of the
turn-extension curves. As can be seen in Fig.~\ref{fig:singmult} the
multi-plectoneme phase accurately describes the experiments, even for very low
salt concentrations (\( 20\, \)mM, data from~\cite{Maffeo:2010,*Brutzer:2010}).

\begin{figure}[htb]
  \centering
  \includegraphics[width=7.0cm]{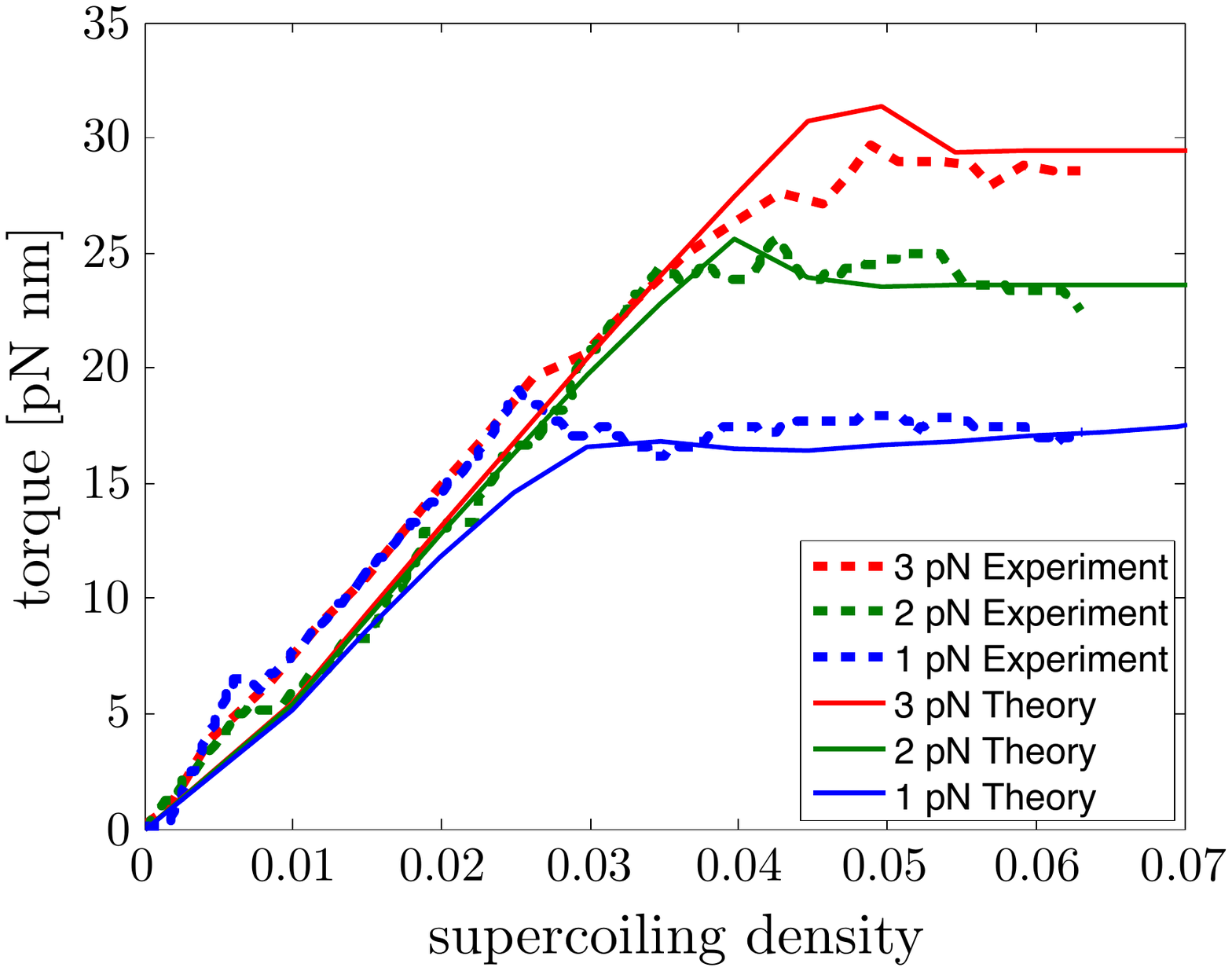}
  \caption{The supercoiling density-torque plots of a $725\,$nm DNA chain at $100\,$mM ionic
      strength for $1\,$pN, $2\,$pN and $3\,$pN tension. Comparison
      between theory and torques measured using a specially crafted cylinder. Data taken from~\cite{Forth:2008}}
  \label{fig:torquef}
\end{figure}

Besides the slope, the multi-plectoneme phase influences the torque after the
transition. In fact, after the transition into the single-plectoneme phase, \(
n \) is transferred at a fixed rate into \( l_p \). This results in a small
bump in the torque at the transition, caused by the use of a number of turns
clamp, followed by a constant plectoneme torque. On the other hand, in the
multi-plectoneme phase the entropic contribution to the free energy ceases to
be linear in \( n \). This explains the difference between torques measured in
optical tweezer experiments~\cite{Forth:2008} and calculated using Maxwell
relations in a magnetic tweezer setup~\cite{Mosconi:2009}. The latter method
is based on the assumption of a constant torque after the transition. However,
for the multi-plectoneme phase our theory predicts a non-constant torque.
In Fig.~\ref{fig:torquem} we show what our
model predicts for the data presented in~\cite{Mosconi:2009}. To facilitate comparison with the original paper, not the linking number, but the supercoiling density is used, which is defined as the ratio of the linking number density to the linking density of the two strands of free DNA. As can be seen in
Fig.~\ref{fig:torquem}, the assumption of constant torque underestimates
the torque difference between the high and low tension curves. The
multi-plectoneme phase prediction, however, correctly reproduces the torque
directly measured in~\cite{Forth:2008}, as is shown in Fig.~\ref{fig:torquef}.

A final consequence of the multi-plectoneme phase is the change in the dynamics of plectonemes in a chain torsionally loaded. Twist mediated plectoneme length
redistribution over the plectonemes makes a fast diffusion of plectonemes possible also in the crowded environment of the plasmoid in bacteria or a dense
chromatin fiber in eukaryotes. The implications for cellular processes from transcription to segregation are significant. 

To conclude: the results of the theory show how the multi-plectoneme phase is
crucial to understand the static and dynamic behavior of DNA under tension and torque.
The torque in fact is much higher than what was previously
computed~\cite{Mosconi:2009}, a fact that could be crucial in the understanding
of the life processes in which DNA is involved.

\begin{acknowledgments}
The authors thank Ralf Seidel for providing us with experimental data and Theo
Odijk and Ralf Seidel for fruitful discussions. We acknowledge discussions with Cees Dekker and 
Marijn van Loenhout.
\end{acknowledgments}

\bibliographystyle{apsrev4-1}
\bibliography{paper}
\end{document}